\documentclass[a4paper,11pt]{article}
\usepackage{pos}

\title{Closure Testing the MSHT PDFs and a First Direct  \\ \vspace*{0.15cm} Comparison to the  Neural Net Approach}

\author*[a]{L.A. Harland-Lang}
\author[a]{R.S. Thorne}
\author[b]{T.Cridge}

\affiliation[a]{Department of Physics and Astronomy, University College London, London, WC1E 6BT, UK }

\affiliation[b]{Deutsches Elektronen-Synchrotron DESY, Notkestr. 85, 22607 Hamburg, Germany }

\emailAdd{robert.thorne@ucl.ac.uk}
\emailAdd{thomas.cridge@desy.de}
\emailAdd{l.harland-lang@ucl.ac.uk}

\abstract{We present a brief overview of the first global closure test of the fixed parameterisation (MSHT) approach to PDF fitting. We find that the default MSHT20 parameterisation can reproduce the features of the input set in such a closure test to well within the textbook uncertainties. This provides strong evidence that parameterisation inflexibility in the MSHT20 fit is not a significant issue in the data region. We also present the first completely like--for--like comparison between two global PDF fits, namely MSHT and NNPDF, where the only difference is guaranteed to be due to the fitting methodology. To achieve this, we present a fit to the NNPDF4.0 data and theory inputs with the MSHT  parameterisation. We find that this gives a moderately, but noticeably, better fit quality than the central NNPDF4.0 fits and that this difference persists at the level of the PDFs and benchmark cross sections. The NNPDF4.0 uncertainties are found to be broadly in line with the MSHT results if a textbook $T^2=1$ tolerance is applied, but to be significantly smaller if a tolerance typical of the MSHT20 fit is applied. }

\FullConference{%
  DIS2024\\
  8-12 April 2024 \\
  Grenoble, France
}

\begin{document}
\hspace*{\fill} DESY-24-123
\maketitle

\section{Introduction}

\noindent Global PDF fits now combine high precision theory, such that NNLO is the standard and now even approximate ${\rm N}^3$LO QCD precision is being accounted for~\cite{McGowan:2022nag,NNPDF:2024nan}, with a wealth of data from multiple experiments and, increasingly, the LHC. Combing these ingredients, we can hope to constrain proton structure with high levels of precision and accuracy. However, as well as differing in the data and the relevant theoretical ingredients that enter the fit, these global PDF analyses also differ rather significantly in the methodologies they apply. Moreover, these methodological effects alone are observed to have a significant impact on the resulting PDFs and, crucially, their uncertainties. 

It is the aim of the study presented here, and described in more detail in~\cite{Harland-Lang:2024kvt}, to begin to address this issue directly. Namely, we  present the first full global closure test of a fixed parameterisation approach to PDF fitting, focusing on the MSHT20~\cite{Bailey:2020ooq} case, and the first completely like--for--like comparison between two global PDF fits, namely MSHT and NNPDF~\cite{NNPDF:2021njg}, where the only difference is guaranteed to be due to the fitting methodology, and PDF parameterisation in particular. We present a brief selection of results relating to these two studies below.

\section{Closure Testing the MSHT Framework}

\noindent This study  makes use of the publicly available \texttt{NNPDF} fitting code~\cite{NNPDF:2021uiq}. By developing a suitable interface to this code, it is possible for us to evaluate the fit quality within the NNPDF framework (i.e. with their theory and choice of dataset, or any subset of it) but for an arbitrary PDF set. Here, we will take this to be parameterised using the fixed basis of Chebyshev polynomials used by the MSHT collaboration. This interface then also allows for a PDF fit (either as a closure test or to real data) to be performed to the NNPDF theory and datasets, but whilst using the MSHT parameterisation of the underlying PDFs at the input scale $Q_0$.

Two types of closure test are performed, namely  `unfluctuated' and `fluctuated',  where the pseudodata are not, and are fluctuated by the  experimental uncertainties, respectively. For both cases we generate pseudodata corresponding to the NNPDF4.0 global dataset, using the central replica of the NNPDF4.0 (perturbative charm) set~\cite{NNPDF:2021njg}. In the unfluctuated test, if the PDF parameterisation in the fit is flexible enough, then a perfect fit (with $\chi^2=0$) is in principle achievable, and any deviations from this will provide an assessment of any lack of flexibility in the  fixed MSHT20 parameterisation. We find a $\chi^2/{\rm N}_{\rm pts}$  of 0.0005, i.e. a very close level of agreement and indeed somewhat lower than that found in the corresponding NNPDF4.0 tests~\cite{DelDebbio:2021whr}. 

\begin{figure}
\begin{center}
\includegraphics[scale=0.55]{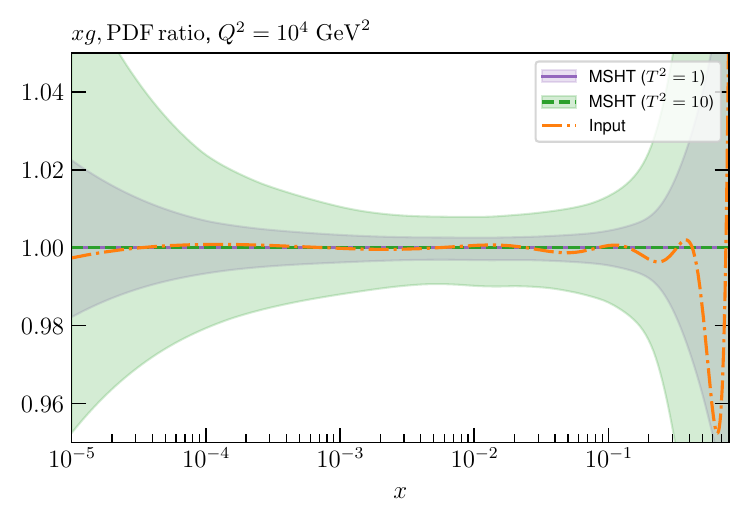}
\includegraphics[scale=0.55]{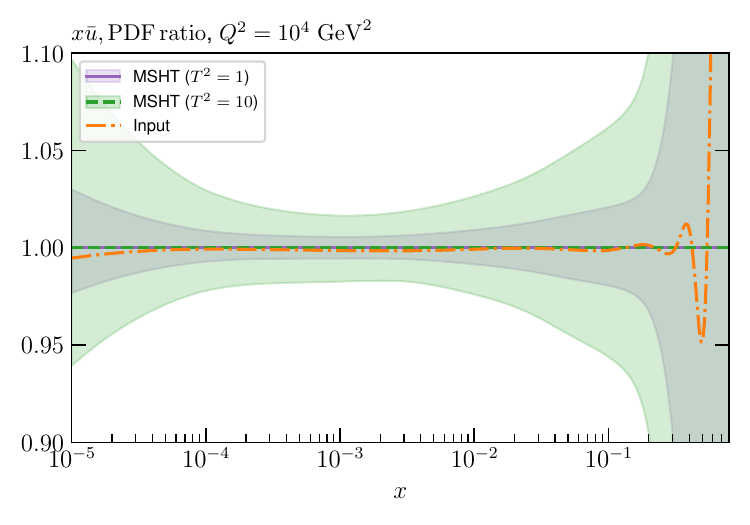}
\caption{\sf \footnotesize{A selection of PDFs at $Q^2=10^4 \, {\rm GeV^2}$  that result from an unfluctuated closure test fit to the NNPDF4.0 dataset,  using the MSHT20 parameterisation. The PDF uncertainties calculated with a $T^2=1$ ($T^2=10$) fixed tolerance are shown in purple (green) and  the NNPDF4.0 (p. charm) input is given by the dashed orange line. Results are shown as a ratio to the MSHT fits.}}
\label{fig:glcl_rat}
\end{center}
\end{figure}

A representative comparison between the fit PDFs and the input is shown in Fig.~\ref{fig:glcl_rat}. To guide this, PDF uncertainties according to the textbook $\Delta \chi^2 = T^2=1$ and the enlarged $T^2=10$ case, similar in size to that taken by MSHT, are indicated. The  agreement between the input and fit  is found to be very good in the data region, at the per mille level, corresponding to  $\sim 10\%$  of the overall $T^2=1$ uncertainty. In the extrapolation region at high $x$ (and in the poorly constrained low to intermediate $x$ valence distributions, not shown) the agreement is less good, as we might expect.

We next consider some results of the fluctuated closure test. We in particular generate 100 pseudodata sets due to the same NNPDF4.0 (p. charm) input set as before, in each case with the pseudodata fluctuated with a different random number seed. Fitting the ensemble of these we can therefore generate a set of 100 fits with which to perform the closure test. While the unfluctuated test assesses the faithfulness of the MSHT fixed parameterisation approach in fitting the central value of the input set, the fluctuated test assesses the faithfulness of the PDF uncertainties that come out of the fit. In particular, if the uncertainties are faithful, then we expect the input PDF set to lie within the 1$\sigma$ uncertainty band of the fitted PDFs with close to 68\% frequency across the  fits.

\begin{figure}
\begin{center}
\includegraphics[scale=0.55]{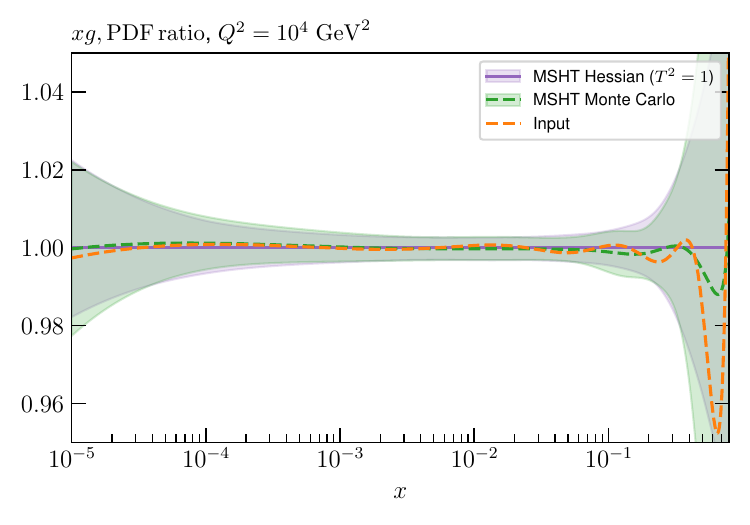}
\includegraphics[scale=0.55]{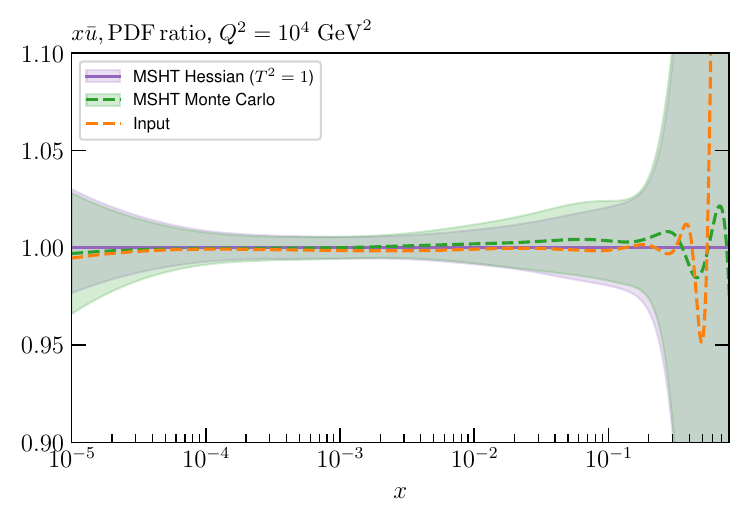}
\caption{\sf \footnotesize{A selection of PDFs at $Q^2=10^4 \, {\rm GeV^2}$  that result from a fluctuated closure test fit to the NNPDF4.0 dataset,  using the MSHT20 parameterisation. The Hessian PDF uncertainties calculated with a $T^2=1$  fixed tolerance that result from the unfluctuated closure test are shown in purple (and are as in Fig.~\ref{fig:glcl_rat}), while the result of fluctuated closure test are shown in green. The NNPDF4.0 (p. charm) input is given by the dashed orange line. Results are shown as a ratio to the MSHT ($T^2=1$) fit.}}
\label{fig:glcl_mcrep}
\end{center}
\end{figure}

A visual indication that this is achieved is shown in Fig.~\ref{fig:glcl_mcrep}, which compares the result of the unfluctuated fit to the 68\% C.L. ensemble of best fits to the 100 fluctuated pseudodatasets. We can we can  see that in the data region the matching between the MC replica and Hessian result is rather close. Combining this with the observation that the Hessian uncertainty on any of the 100 fluctuated fits is very stable, and  closely matches this $T^2=1$ Hessian uncertainty on the unfluctuated fit, is then sufficient to demonstrate that the closure test has been passed at the level of each individual PDF and at any given $x$ point in the data region. In other words, each of these 100 fits will be consistent within its 1$\sigma$ Hessian uncertainty with the input set with close to 68\% frequency. 

\section{A Direct Comparison to NNPDF4.0}

\noindent We now fit the real data corresponding to the same NNPDF4.0 dataset; we will refer to these as `MSHT fits' for brevity. We recall in particular that both the data in the fit, but also the theory settings, are completely  identical to those used in the corresponding NNPDF4.0 fit. The comparison is therefore completely like--for--like at the level of the full global PDF fit, with the only difference being due to the PDF parameterisation. The global fit qualities are shown in Table~\ref{tab:chi2}, for both perturbative and fitted charm. In the latter case the charm quark PDF is freely parameterised using the same basis of Chebyshev polynomials taken for other PDFs in the MSHT analysis, and this represents the first model independent fitted charm determination with a fixed parameterisation.

\begin{table}
\begin{center}
  \scriptsize
  \centering
   \renewcommand{\arraystretch}{1.4}
\begin{tabular}{rccc}\hline 
&NNPDF4.0 pch&MSHT fit&MSHT fit (w positivity)\\
\hline
 Perturbative Charm (4626)  & 5928.3 (1.282)& 5736.7 (1.240) &5837.8 (1.262)\\
\hline
 Fitted Charm, (4666)  &  5692.1 (1.233)&  5645.2 (1.222) &5651.0 (1.224)\\
\hline
\hline
\end{tabular}
\end{center}
\caption{\sf \footnotesize{Global $\chi^2$ ($t_0$ definition) values for the NNPDF4.0 fit and the MSHT fits to the NNPDF dataset/theory settings, with both perturbative and fitted charm, and with and without positivity imposed for the MSHT fits.}}
\label{tab:chi2}
\end{table}

We can see that, despite the nominally larger flexibility  of the NNPDF4.0 fit, the MSHT fit produces a moderately, but noticeably, better fit quality than the central NNPDF results, both with perturbative and fitted charm. In the former case, this can in part be explained by positivity of the PDFs imposed by NNPDF, in particular of the gluon at low $x$ and $Q^2$, and which is not by default imposed in the MSHT fit. Accounting for this leads to some deterioration in the perturbative charm fit, but with a result that remains better than the  NNPDF value. In~\cite{Harland-Lang:2024kvt} the question of why this occurs is considered in detail, and the possibility that the MSHT parameterisation might lead to overfitting closely examined, but no evidence of this found. Conversely, it is demonstrated that if a more restrictive parameterisation is taken, then the closure tests described in the previous section are not passed successfully, and a poorer description in the full fit is arrived at.

\begin{figure}
\begin{center}
\includegraphics[scale=0.55]{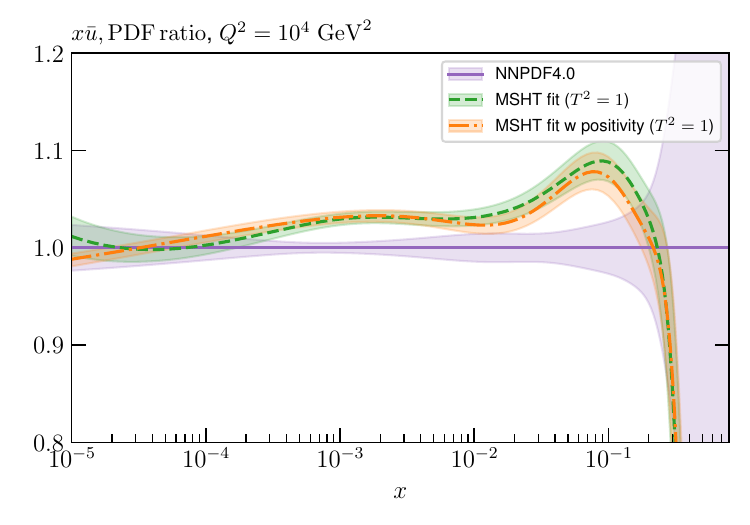}
\includegraphics[scale=0.55]{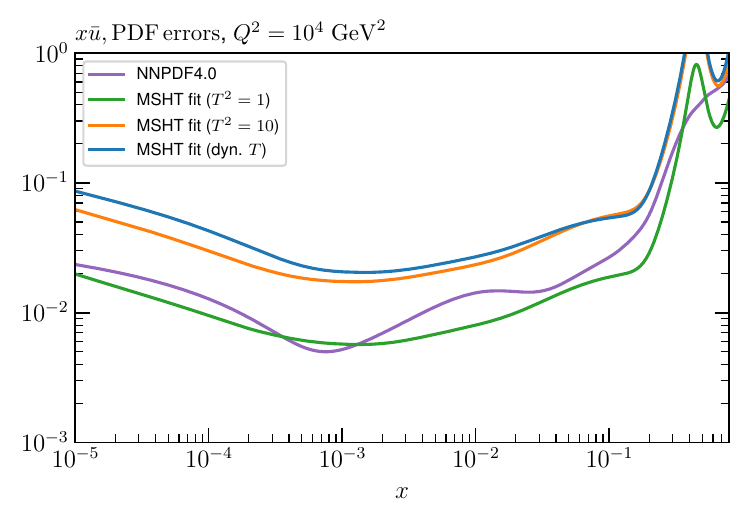}
\caption{\sf \footnotesize{ The $\overline{u}$ at $Q^2=10^4 \, {\rm GeV^2}$  that results from a  global PDF fit to the NNPDF4.0 dataset/theory (fitted charm) setting, but using the MSHT20 parameterisation. Results with and without a positivity constraint applied are shown. In the left plot the ratio to the NNPDF4.0 set is shown, with PDF uncertainties for the MSHT fits corresponding to a fixed $T^2=1$ tolerance. In the right plot the corresponding PDF uncertainties are shown, for different fixed and dynamic tolerances in the MSHT case.}}
\label{fig:pdfcomp}
\end{center}
\end{figure}

We also find some significant differences in the  PDFs, such that those due to the NNPDF4.0 releases are not always compatible with these MSHT fits within the nominal NNPDF uncertainties. These differences tend to be larger in the perturbative charm fits, but are also present with fitted charm. An example of this is shown in Fig.~\ref{fig:pdfcomp}. In the left plot the ratio to the NNPDF result is shown, with $T^2=1$ uncertainties taken in the MSHT results for ease of comparison, and a clear difference is observed. In the right plot the PDF uncertainties are given, and we can see that the NNPDF uncertainty is similar in size to the MSHT case with $T^2=1$, but smaller than that with $T^2=10$. Similar results are found for other PDFs,  notably with respect to the quark flavour decomposition. 

The PDF uncertainty comparison in particular demonstrates that the NNPDF4.0 fit produces PDF uncertainties that are rather more closely in line with those of the fixed MSHT20 parameterisation if the textbook $T^2=1$ criterion is applied. Given the default MSHT20 approach applies a larger tolerance of $T^2 \sim 10$, this  points to an inherent inconsistency between the approaches, although as discussed in more detail in~\cite{Harland-Lang:2024kvt} there are very good reasons for including an enlarged, $T^2>1$, tolerance in global PDF fits. Evidence for this is demonstrated in this work at the level of closure tests, where it is shown that if data/theory inconsistencies are included in the closure test, then these do not lead to any increase in the $T^2=1$ PDF uncertainty.

In summary, we have presented a selection of results from the first global closure test of the fixed parameterisation (MSHT) approach to PDF fitting and first completely like--for--like comparison between two global PDF fits, namely MSHT and NNPDF. We find that the default MSHT20 parameterisation can reproduce the features of the input set in such a closure test well, while 
for the full fit we find a moderately, but noticeably, better fit quality than the central NNPDF4.0 fits.

\end{document}